\documentclass[12pt,preprint]{aastex}

\begin{document}

\title{MICROLENSING OPTICAL DEPTH REVISITED WITH RECENT STAR COUNTS}
\lefthead{Ryu et al.} \righthead{MICROLENSING OPTICAL DEPTH
REVISITED WITH RECENT STAR COUNTS}


\author{Yoon-Hyun Ryu, Heon-Young Chang\footnote{corresponding author},
Myeong-Gu Park,}

\affil{Department of Astronomy and Atmospheric Sciences, Kyungpook
National University, Daegu 702-701, Korea; yhryu@knu.ac.kr,
hyc@knu.ac.kr, mgp@knu.ac.kr}

\and

\author{Ki-Won Lee}
\affil{Korea Astronomy and Space Science Institute, Daejeon 305-348,
Korea; kwlee@kasi.re.kr}


\begin{abstract}
More reliable constraints on the microlensing optical depth comes
from a better understanding of the Galactic model. Based on
well-constrained Galactic bulge and disk models constructed from
survey observations, such as, $\it{HST}$, 2MASS, and SDSS, we
calculate the microlensing optical depths toward the Galactic bulge
fields, and compare them with recent results of microlensing
surveys. We test $\chi^2$ statistics of microlensing optical depths
expected from those models, as well as previously proposed models,
using two types of data: optical depth map in $(l, b)$ and averaged
optical depth over the Galactic longitude $l$ as a function of the
latitude $b$. From this analysis, we find that the Galactic bulge
models of 2MASS, Han $\&$ Gould (2003), and G2 of Stanek et al.
(1997) show a good agreement with the microlensing optical depth
profiles for all the microlensing observations, compared with E2 of
Stanek et al. (1997). We find, on the other hand, that models
involving an SDSS disk model produce relatively higher $\chi^2$
values. It should be noted that modeled microlensing optical depths
diverge in the low Galactic latitude, $|b| \la 2 \arcdeg$.
Therefore, we suggest the microlensing observation toward much
closer to central regions of the Galaxy to further test the proposed
Galactic models,  if it is more technically feasible than waiting
for large data set of microlensing events.
\end{abstract}


\section{INTRODUCTION}

Microlensing survey was originally proposed  as a tool for detecting
massive astronomical compact halo objects (MACHOs) in the Galactic
halo \citep{pac86}. Searches for the microlensing events towards the
Magellanic Clouds by the MACHO \citep{alc93} and EROS groups
\citep{aub93} have placed constraints on the  fraction of the
Galactic MACHO populations. Searches for dark objects in the halo of
M31 have also been performed (AGAPE, Ansari et al. 1999; WeCAPP,
Riffeser et al. 2003; MEGA, de Jong et al. 2004; VATT-Columbia,
Uglesich et al. 2004; POINT-AGAPE, Calchi Novati et al. 2005;
Angstrom, Kerins et al. 2006). In addition, microlensing has proven
itself as a powerful tool in constraining the distribution of faint
stellar objects in the Milky Way
\citep{kir94,han95,han03,woo05,woo07,cal08}. For instance,
\citet{han95} examined theoretical Galactic models by constructing a
microlensing map. They explored a triaxial bar-shaped bulge model,
which was based on by $\it{COBE}$-DIRBE multiwavelength observations
of the Galactic bulge \citep{wei94}, and demonstrated that
observationally determining the microlensing optical depth may
provide supplementary information on the Galactic mass distribution.

Unfortunately, however, reality was not so simple. Early
measurements of the microlensing optical depth towards the Galactic
bulge were significantly higher than predictions, suggesting both
observational and theoretical sides make efforts to reconcile.
Measurements at Baade's window by OGLE from 9 microlensing events
\citep{uda94} and at $(l,b)=(2\fdg55, -3\fdg64)$ by MACHO from 13
events of clump giant sources \citep{alc97a} have yielded
$\tau=3.3\pm1.2\times10^{-6}$ and
$\tau=3.9^{+1.8}_{-1.2}\times10^{-6}$, respectively. \citet{alc97a}
pointed out a possibility of a systematic bias such as blending
effect in the microlensing optical depth measurement to explain
those high optical depth values in the sense that bulge fields are
high dense regions. \citet{pop01} suggested that the bias due to the
blending can be avoided by using the events only with bright source
stars such as red clump giant stars. Such stars are identified by
their well-defined position in the color-magnitude diagram. This
position ensures that they are most likely in the Galactic Bulge.
Their large flux also makes blending problems relatively
unimportant. Measurements of $\tau$ using red clump giants have
tended to reduce discrepancies with models. Based on 16 events, the
EROS-2 collaboration \citep{afo03} gave the microlensing optical
depth of $0.94\pm0.29\times 10^{-6}$ at $ (l,b) =(2\fdg5 ,-4\fdg0)$.
The MACHO group calculated the microlensing optical depth toward the
Galactic bulge using their seven-year survey data \citep{pop05}. The
measurement of $\tau=2.17^{+0.47}_{-0.38}\times10^{-6}$ at
$(l,b)=(1\fdg50,-2\fdg68)$ was obtained from 42 microlensing events
detected during the monitoring of about $7.4\times10^5$ clump giant
sources covering 4.5 deg$^2$. More recently, OGLE-II presented the
measurement of the microlensing optical depth toward the Galactic
bulge based on recent four-year survey data \citep{sum06}. Using the
sample of 32 microlensing events in 20 bulge fields covering $\sim5$
deg$^2$, they found $\tau=2.55^{+0.57}_{-0.46}\times10^{-6}$ at
$(l,b)=(1\fdg16,-2\fdg75)$. Efforts in a theoretical side also
refined the microlensing optical depth estimate by using more
sophisticated models based on observational results from wide field
surveys. Theoretical estimates involving a bar in the bulge oriented
along the line of sight approach to a value in the range of
$0.8\times10^{-6}$ to $2.0\times10^{-6}$ \citep[see
e.g.][]{pac94,zha95,zha96a,bin00,han03}.

The optical depth in microlensing is defined as the probability that
a source star is being placed within the Einstein radius of a
foreground lens star. If bulge stars are distributed over a distance
$d$ from the Earth, the microlensing optical depth is given by
\begin{equation}
\tau=\frac{4\pi G}{c^2}\frac{\int^{d}_0
\left[\int^{D_s}_0dD_l\;\rho(D_l)\;D\right]dD_s\;D_s^2\;n(D_s)}
{\int^{d}_0dD_s\;D_s^2\;n(D_s)},
\end{equation}
where $n(D_s)$ is the number density of source stars, $\rho(D_l)$ is
the mass density of lenses along the line of sight, $D_l$, $D_s$ and
$D_{ls}$ are the distances from the observer to the lens and source,
and the distance from the lens to the source, respectively, and $D$
is $D_lD_{ls}/D_s$.

Given that the microlensing optical depth is subject to an
underlying mass distribution model, it is crucial to employ a mass
model, which can be constructed from independent observations, e.g.,
star counts. In fact, the stellar contents of the disk and the bulge
have been measured by survey projects and modeled to estimate the
optical depth. For instance, star counts from {\it Hubble Space
Telescope} ({\it HST}) provided constraints on faint stars down to
the hydrogen burning limit in the disk and 0.15 $M_{\odot}$ in the
bulge \citep{hol98,zoc00,zhe01}. Using those observational results,
one may put relatively stronger constraints on the mass distribution
in terms of the microlensing optical depth \citep[e.g.][]{han03}.
Results from large-scale sky surveys such as 2MASS \citep{skr97} and
SDSS \citep{yor00} are also available in the near-infrared and
optical wavelength bands. On the basis of those results, new models
of the Galactic bulge and disk have been suggested
\citep{lop05,jur08}. In the present paper, we calculate the
microlensing optical depth with Galactic bulge and disk models
recently by various models based on recent survey observations as
well as with those studied earlier. We then compare the optical
depths based on models with the results of MACHO, OGLE-II, and
EROS-2 surveys.

The paper is organized as follows. We begin with a brief description
of models of the Galactic bulge and the disk in $\S$ 2. We present
expected microlensing optical depths from various models and compare
those with observed optical depths in $\S$ 3. Finally, we conclude
with a summary of our results and discussion.

\section{GALACTIC BULGE AND DISK MODELS}

\subsection{BULGE MODEL}

We calculate theoretical optical depths based on 5 different mass
distribution models. To begin with, we follow \citet{lop04,lop05},
who have suggested two bulge models, i.e., triaxial and boxy models,
on the basis of the number density from 2MASS star counts
\citep{skr97}. That is, our first and second models are
\begin{equation}
\rho_{\rm{triaxial}}(x,y,z)=\rho_{\rm{T}}\exp\left(-\frac{t_1}{740\;\rm{pc}}\right),
\end{equation}
and
\begin{equation}
\rho_{\rm{boxy}}(x,y,z)=\rho_{\rm{B}}\exp\left(-\frac{0.866\;t_2}{740\;\rm{pc}}\right),
\end{equation}
respectively, where
\begin{equation}
t_1=\left[x^2+\left(\frac{y}{0.49}\right)^2
+\left(\frac{z}{0.37}\right)^2\right]^{1/2},
\end{equation}
\begin{equation}
t_2=\left[x^4+\left(\frac{y}{0.49}\right)^4+
\left(\frac{z}{0.37}\right)^4\right]^{1/4}.
\end{equation}
The normalization in the triaxial and boxy bulge models are
$\rho_{\rm{T}}=9.9$ $M_\odot/\rm{pc}^3$ and $\rho_{\rm{B}}=4.67$
$M_\odot/\rm{pc}^3$, respectively. Following \citet{cal08} we assume
the total bulge mass of $1.5\times10^{10}$ $M_\odot$ within 2.5 kpc
from the Galactic center. Note that this value is in a range of
generally accepted bulge mass of the Galaxy $1-2\times10^{10}$
$M_\odot$ \citep[e.g.][]{blu95,zha96b,deh98}. For both bulge models,
the longest axis is inclined by $29\arcdeg$ with respect to the line
of sight.

We also employ three widely-used bulge models. Our third bulge model
(HG03) is one studied by \citet{han03}, which is favored by an
analysis of the \emph{COBE} DIRBE observations \citep{dwe95}.
\citet{han03} normalized the model using the $\it{HST}$ star count.
For the fourth model, we adopt G2 model, which can be found in
\citet{sta97}, given as
\begin{equation}
\rho_{\rm{G2}}(x,y,z)=\rho_0 \exp\left(-\frac{r^2_s}{2}\right),
\end{equation}
where
\begin{equation}
r_s=\left[\left\{\left(\frac{x}{x_0}\right)^2+\left(\frac{y}{y_0}\right)^2\right\}^2
+\left(\frac{z}{z_0}\right)^4\right]^{1/4}.
\end{equation}
Following \citet{sta97} we take $x_0=1239~\mathrm{pc}$ with the axis
ratio of $x_0:y_0:z_0 = 2.8:1.4:1$ and an inclination angle $\alpha
= 24\fdg9$. The last model is E2 model of \citet{sta97}. The density
distribution is given in the form of
\begin{equation}
\rho_{\rm{E2}}(x,y,z)=\rho_0 \exp(-r),
\end{equation}
where
\begin{equation}
r=\left[\left(\frac{x}{x_0}\right)^2+\left(\frac{y}{y_0}\right)^2+\left(\frac{z}{z_0}\right)^2\right]^{1/2}.
\end{equation}
Here we take $x_0=897~\mathrm{pc}$ with the axis ratio of
$x_0:y_0:z_0 = 3.6:1.5:1$ and an inclination angle of the bulge
major axis with respect to the line of sight $\alpha = 23\fdg8$. For
HG03, G2, and E2 bulge models, we follow the bulge mass
normalizations of \citet{cal08} and \citet{han03}. We set the
distance to the Galactic center as 8.0 kpc throughout our analysis.

\subsection{DISK MODEL}

For 2MASS bulge models we employ two disk models  (disk 1 model and
disk 2 model) derived from 2MASS star counts \citep{lop04,lop05}.
For bulge models of HG03, G2, and E2 we adopt the disk model of
\citet{zhe01}, whose density profile is given as a $\mathrm{sech}^2$
(exponential) function for the thin (thick) components. We normalize
it with the disk mass density of the Solar neighborhood,
$\rho_\odot=0.05$ $M_\odot/\rm{pc}^3$ \citep[cf.][]{han03}. In
addition, we also investigate the SDSS disk model with respect to
all the bulge models we consider in the present paper. \citet{jur08}
estimate the three-dimensional number density distribution of the
Galactic disk and halo using the photometric parallax method on the
SDSS data. We use their disk models consisting of thin and thick
exponential disk with local thick-to-thin disk  normalization
$\rho_{{\rm thick}}(R_\odot)/\rho_{{\rm thin}}(R_\odot)=12\%$.
Again,  the local number density is normalized to
$\rho(R_\odot)=0.05$ $M_\odot/\rm{pc}^3$.

\section{RESULTS}

In Figures 1 and 2, we show the map of the optical depth difference
normalized by the observational error $\sigma_{\tau}$ for each $(l,
b)$ field, which is given by
$\epsilon\equiv(\tau_{o}-\tau_{m})/\sigma_{\tau}$, where $\tau_{o}$
and $\tau_{m}$ are the microlensing optical depths in a given field
from observation and model, respectively. We compare the
microlensing optical depth observed from MACHO and OGLE-II with
those calculated from models in field by field. Figures 1 and 2
result from MACHO and OGLE-II, respectively. In the map, the fields
of negative $\epsilon$ and positive $\epsilon$ are shown in blue and
red, respectively. For those of $\tau_{o}=0.0$, fields are marked by
black. Dotted curves in each panel represent the microlensing
optical depth contour of a model in units of $10^{-6}$. Each panel
of Figures 1 and 2 results from different mass models indicated in
the lower left corner of the panel.  Table 1 summarize how to pair a
bulge model and a disk model. For instance,  2MT1 and 2MB1 represent
combinations of 2MASS triaxial bulge model and the 2MASS disk 1
model, 2MASS boxy bulge model and the 2MASS disk 1 model,
respectively. In case of 2MT2 and 2MB2, the 2MASS disk 1 model is
replaced by the 2MASS disk 2 model for a given bulge model. When the
SDSS disk model is substituted, it is indicated as shown: HG03, G2,
and E2 stand for cases of HG03, G2, and E2 bulge models and the disk
model of \citet{zhe01}.

To quantify the difference of the microlensing optical depth, we
calculate $\chi^2_{l,b}$, defined as
\begin{equation}\label{chi}
\chi^2_{l,b}\equiv\sum^N_{i=1}\frac{(\tau_{o,i}-\tau_{m,i})^2}{\sigma_{o,i}^2},
\end{equation}
where $\tau_{o,i}$, $\tau_{m,i}$ and $\sigma_{o,i}$ are the
microlensing optical depths from observation and model, and the
observational error in each field, respectively. Here summations are
repeated over  14 OGLE-II fields and 19 MACHO fields excluding
fields of $\tau_o=0$. We summarize results in Table 2. As seen, one
finds that 2MASS, HG03, and G2 models agree well with the
microlensing optical depth distribution for MACHO and OGLE-II when
disk models of 2MASS and \citet{zhe01} are used. It should be noted
that models that include the SDSS disk model produce relatively
higher $\chi^2_{l,b}$ values. We note that the microlensing optical
depth from MACHO has much higher $\chi^2_{l,b}$ values than OGLE-II
since the microlensing optical depth values in four MACHO fields
(176, 113,109, and 105) are 3$\sigma$ below with respect to the
microlensing optical depth values expected from all models in those
fields. Therefore, it should be pointed out that comparing two
columns of OGLE-II and MACHO is meaningless. We repeat calculation
of $\chi^2$ without those four fields and show results in
parentheses for comparison in Table 2. Conclusion remains same. The
reason why four of the MACHO fields deviate out of 3 $\sigma$ could
be either that the models are too smooth to explain details of
Galactic structures while there could be components affecting more
optically than dynamically or that  observational contaminations
like the blending effect may play a role. For instance, OGLE-II
investigates the blending effect even in bright events and points
out that many bright microlensing events include heavily blended
events \citep[e.g.][]{uda94,alc97b,smi07}.

In Figure 3, we also show microlensing optical depths averaged over
the Galactic longitude as a function of the Galactic latitude $b$.
In the left panel, averages of microlensing optical depths measured
by OGLE-II, MACHO, and EROS-2 are shown with their error bars.
OGLE-II, MACHO, and EROS-2 data are plotted with filled circles,
squares, and triangles, respectively. Best fits are superposed on
top of observed data. Different line types represent different
microlensing experiments: continuous, short dashed and long dashed
lines stand for best-fit curves of OGLE-II, MACHO, and EROS-2,
respectively. Analytic functions of those fits estimated by OGLE-II,
MACHO, and EROS-2 are given as
$\tau_{\rm{OGLE-II}}=[(4.48\pm2.37)+(0.78\pm0.84)\times
b]\times10^{-6}$ \citep{sum06},
$\tau_{\rm{MACHO}}=[5.2+(1.06\pm0.71)\times b]\times10^{-6}$
\citep{pop05}, and
$\tau_{\rm{EROS-2}}=(1.62\pm0.23)\exp[-(0.43\pm0.16)(|b|-3
\arcdeg)]\times10^{-6}$ \citep{ham06}, respectively. In the right
panel, microlensing optical depths toward the Galactic bulge
expected from models described in the previous section are shown as
a function of the Galactic latitude  $b$ with thin curves. Dashed
curve represents the calculated microlensing optical depth resulted
from the 2MASS triaxial bulge model and the 2MASS disk 2 model
(2MT2). Dotted curve represents the calculated microlensing optical
depth resulted from the 2MASS boxy bulge model and the 2MASS disk 2
model (2MB2). Long dashed, dot-long-dashed, and dot-short-dashed
curves represent HG03, G2, and E2 models, respectively. For
comparison, we plot best fits obtained by different microlensing
experiments with thick curves, which are already shown in the left
panel. Continuous, short dashed and long dashed curves correspond to
results of OGLE-II, MACHO, and EROS-2, respectively, as in the left
panel. Models shown in the plot in general seem compatible with all
the microlensing observations within the observational
uncertainties. It is interesting to note that deviations among model
fits become larger as the Galactic latitude decreases. To evaluate
more quantitatively we calculate $\chi^2_b$ in each $b$ using
equation (10), and list results in Table 3. As seen previously in
Table 2, 2MASS, HG03, and G2 result in smaller $\chi^2_b$ in all the
microlensing experiments. In Table 3 we show $\chi^2_b$ of all
cases, including those not shown in the right panel of Figure 3. One
may find again that when SDSS disk model is used fits become poorer.

\section{SUMMARY AND DISCUSSIONS}

By comparing the microlensing optical depth map constructed by a
mass model with the observed microlensing optical depth one may put
a tight constraint on the Galactic model. We have estimated the
microlensing optical depths toward the Galactic bulge using the
Galactic bulge and disk models constructed from survey observations,
and compared them with the sample of microlensing events observed
towards the Galactic bulge with red clump giant sources reported by
the MACHO, OGLE-II and EROS-2 collaborations. From the analysis we
carried out, we show that model estimates are in general compatible
with microlensing observations. According to calculated $\chi^2$ we
find that 2MASS, HG03, and G2 models reproduce the microlensing
optical depth distribution from observations well. We also find that
for those well-fit models contribution to the microlensing optical
depth of the disk components is not crucial. For example, for 2MASS
cases two disk models do not make any difference in $\chi^2$. It is,
however, interesting to note that the analyses of SDSS disk models
give us somewhat different results. That is, models including SDSS
disk model produce relatively higher $\chi^2$ values than others do.

It should be pointed out that current observational data of both the
microlensing optical depth in two dimensional fields and the
averaged optical depth over $l$ are yet statistically insufficient
to rule out conclusively any specific models we have considered
since the number of observed microlensing events from monitoring
RCGs in recent microlensing surveys is still small. On the other
hands, we note that the difference of the average optical depth
among models increases as closer to the Galactic center. Therefore,
we suggest the microlensing observation toward closer to the
Galactic center regions if possible may be more effective in
constraining the Galactic model than extending microlensing searches
to detect more microlensing events.

\acknowledgements We thank the anonymous referee for useful comments
which improve our presentation. This work is the result of research
activities at the Astrophysical Research Center for the Structure
and Evolution of the Cosmos (ARCSEC) supported by the Korea Science
\& Engineering Foundation (KOSEF). HYC was supported by the Korea
Research Foundation Grant funded by the Korean Government (MOEHRD,
Basic Research Promotion Fund) (KRF-2006-311-C00072).

\clearpage \linespread {1.0}

\begin{deluxetable}{lccccc}
\tablecolumns{5} \tablewidth{0pc} \tablecaption{Bulge and disk
models} \tablehead{ \colhead{ } & \colhead{2MASS disk 1}&
\colhead{2MASS disk 2}& \colhead{Zheng disk}& \colhead{SDSS disk}}
\startdata
  2MASS triaxial bulge  &2MT1 &2MT2&      & 2MT+SDSS\\
  2MASS boxy bulge       &2MB1 &2MB2&      & 2MB+SDSS \\
  Han \& Gould (2003)         &     &    &HG03  &HG03+SDSS \\
  G2 of Stanek et al. (1997)   &     &    &G2    &G2+SDSS \\
  E2 of Stanek et al. (1997)   &     &    &E2    &E2+SDSS \\

\enddata

\end{deluxetable}

\begin{deluxetable}{lcc}
\tablecolumns{3} \tablewidth{0pc} \tablecaption{The chi-square
values of microlensing optical depths between observations and model
estimates in two dimensional microlensing optical depth map.}
\tablehead{ \colhead{}    & \multicolumn{2}{c}{$\chi^2_{l,b}$}\\
\cline{2-3}  \\
\colhead{Model} & \colhead{OGLE-II}& \colhead{MACHO}} \startdata
 2MT1      &  13.73  & 122.23 (27.20)\\
 2MT2      &  13.57  & 122.31 (27.51)\\
 2MT+SDSS  &  19.23  & 178.20 (38.76)\\
 2MB1      &  15.24  & 143.84 (34.55)\\
 2MB2      &  14.91  & 142.55 (34.63)\\
 2MB+SDSS  &  21.84  & 208.89 (49.50)\\
 HG03      &  14.90  & 121.72 (24.92)\\
 HG03+SDSS &  17.68  & 156.83 (32.77)\\
 G2        &  16.81  & 159.30 (35.15)\\
 G2+SDSS   &  20.13  & 202.00 (46.15)\\
 E2        &  23.66  & 200.51 (41.89)\\
 E2+SDSS   &  28.05  & 248.51 (54.15)\\

\enddata
\tablecomments{The number of fields used in the summation is 14 and
19 for OGLE-II and MACHO, respectively. For MACHO, chi-square values
excluding 4 fields whose errors are greater than 3 $\sigma$ are also
shown in parentheses for comparison.}
\end{deluxetable}

\begin{deluxetable}{lrrr}
\tablecolumns{4} \tablewidth{0pc} \tablecaption{The chi-square
values of averaged optical depths as a funciton of $b$.}
\tablehead{ \colhead{}    & \multicolumn{3}{c}{$\chi^2_{b}$}\\
\cline{2-4}  \\
\colhead{Model} & \colhead{OGLE-II}& \colhead{MACHO}&
\colhead{EROS-2}}
\startdata
 2MT1      & 2.27  & 1.66    &  10.71\\
 2MT2      & 2.26  & 1.76    &  10.57\\
 2MT+SDSS  & 3.48  & 6.94    &  22.03\\
 2MB1      & 2.76  & 4.41    &  18.41\\
 2MB2      & 2.73  & 4.35    &  18.00\\
 2MB+SDSS  & 4.81  & 13.41   &  35.59\\
 HG03      & 2.41  & 1.76    &  10.83\\
 HG03+SDSS & 3.13  & 5.20    &  17.46\\
 G2        & 3.30  & 6.25    &  21.11\\
 G2+SDSS   & 4.54  & 12.05   &  31.55\\
 E2        & 4.94  & 13.42   &  28.00\\
 E2+SDSS   & 6.60  & 21.26   &  39.22\\

\enddata

\end{deluxetable}

\begin{figure}
    \epsscale{1.0} \plotone{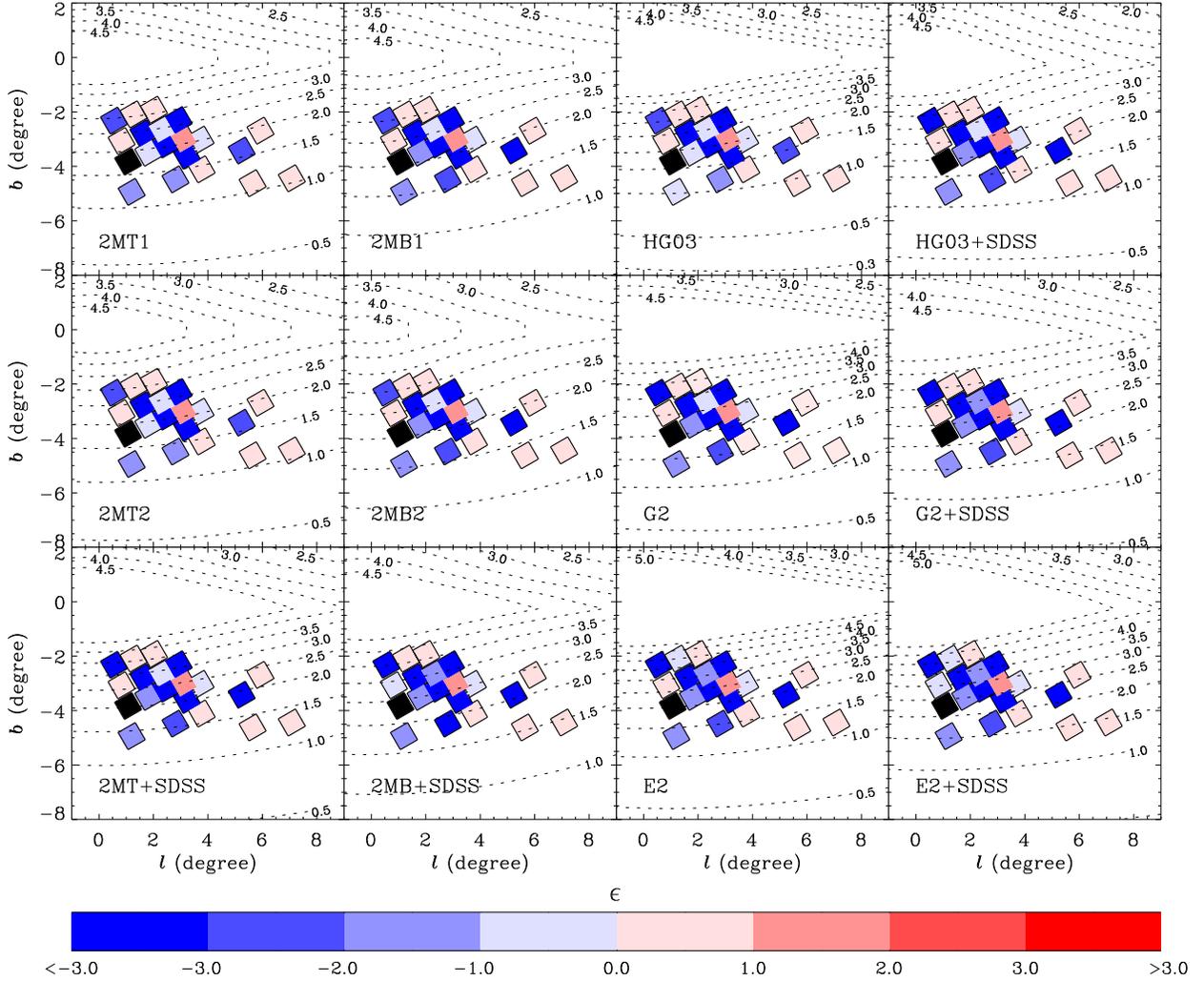}
    \caption{Relative difference of microlensing optical
    depths in $10^{-6}$ between MACHO results and models,
    normalized by the
observational error $\sigma_{\tau}$,  as a function
    of $(l,b)$.
    Note that the fields of negative $\epsilon$ and positive $\epsilon$
are shown in blue and red, respectively. For those of
$\tau_{o}=0.0$, fields are marked by black.    Dotted curves in each
panel represent the microlensing
    optical depth contour of a model in units of $10^{-6}$. Each panel
    results from different mass models indicated in the lower left
    corner of the panel. See the text for the definition of
    the error-normalized difference
    and for model descriptions. }
\end{figure}

\begin{figure}
    \epsscale{1.0} \plotone{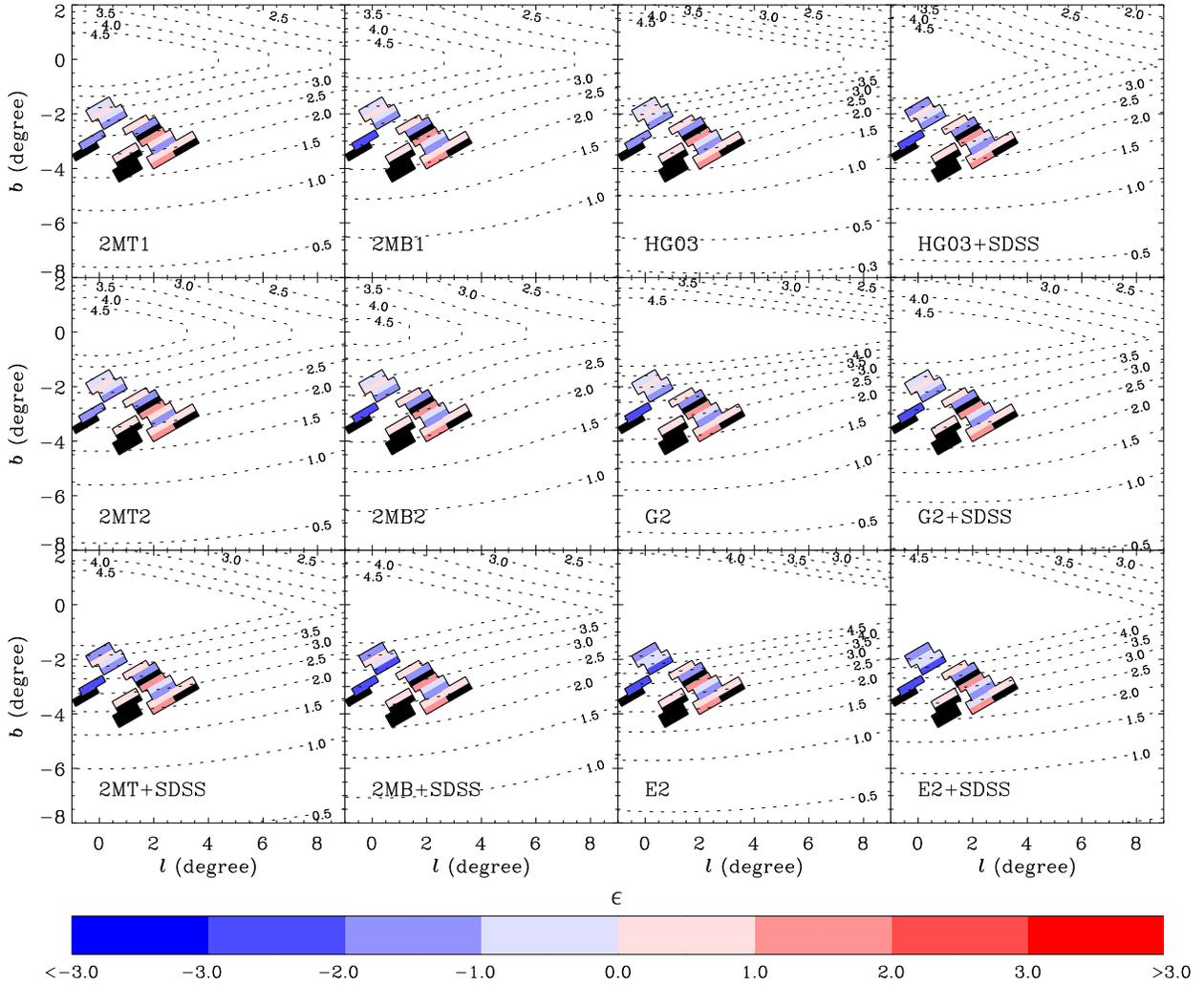}
    \caption{Same plots as Figure 1, but showing results from OGLE-II.}
\end{figure}

\begin{figure}
    \epsscale{1.0} \plottwo{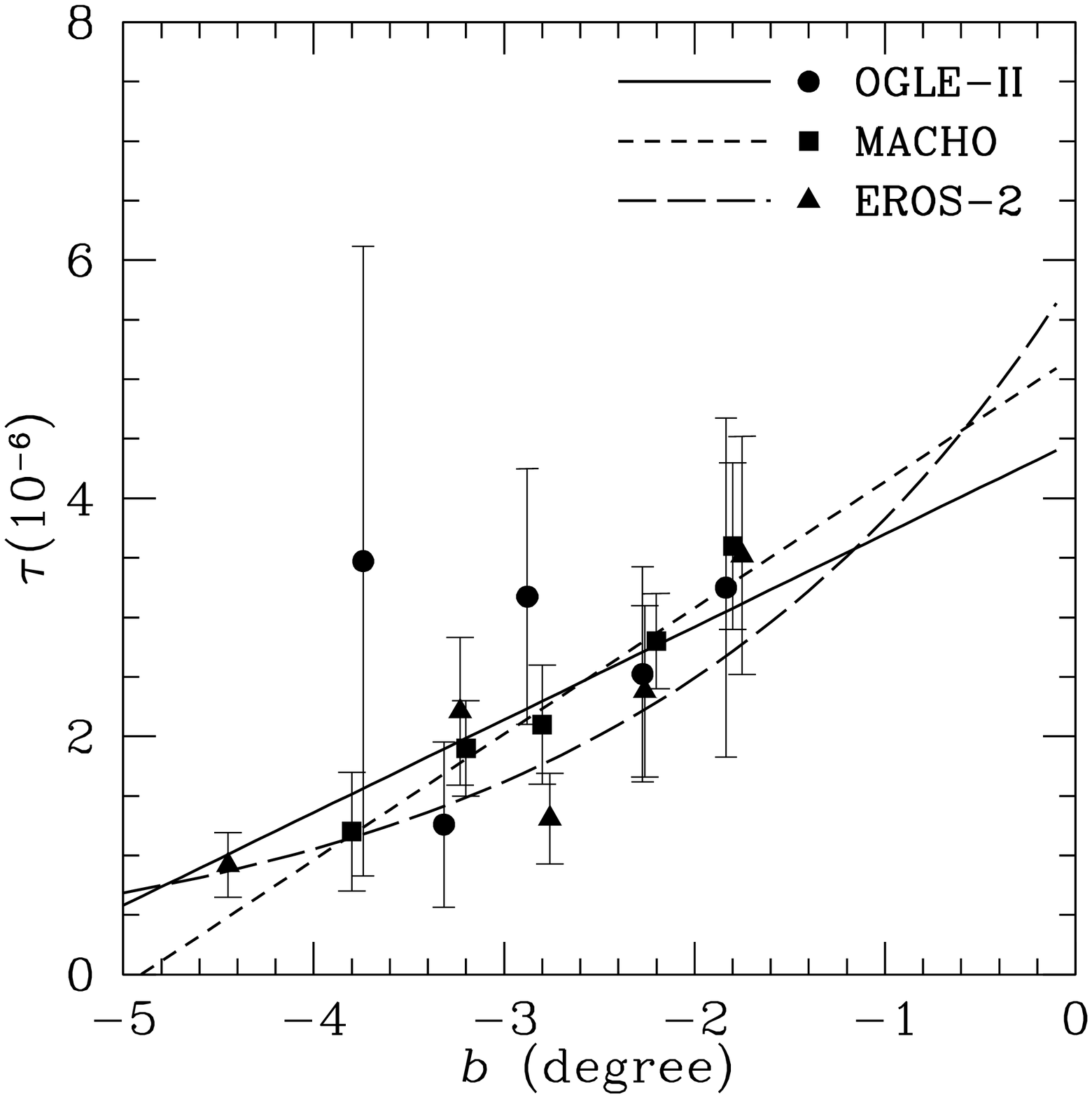}{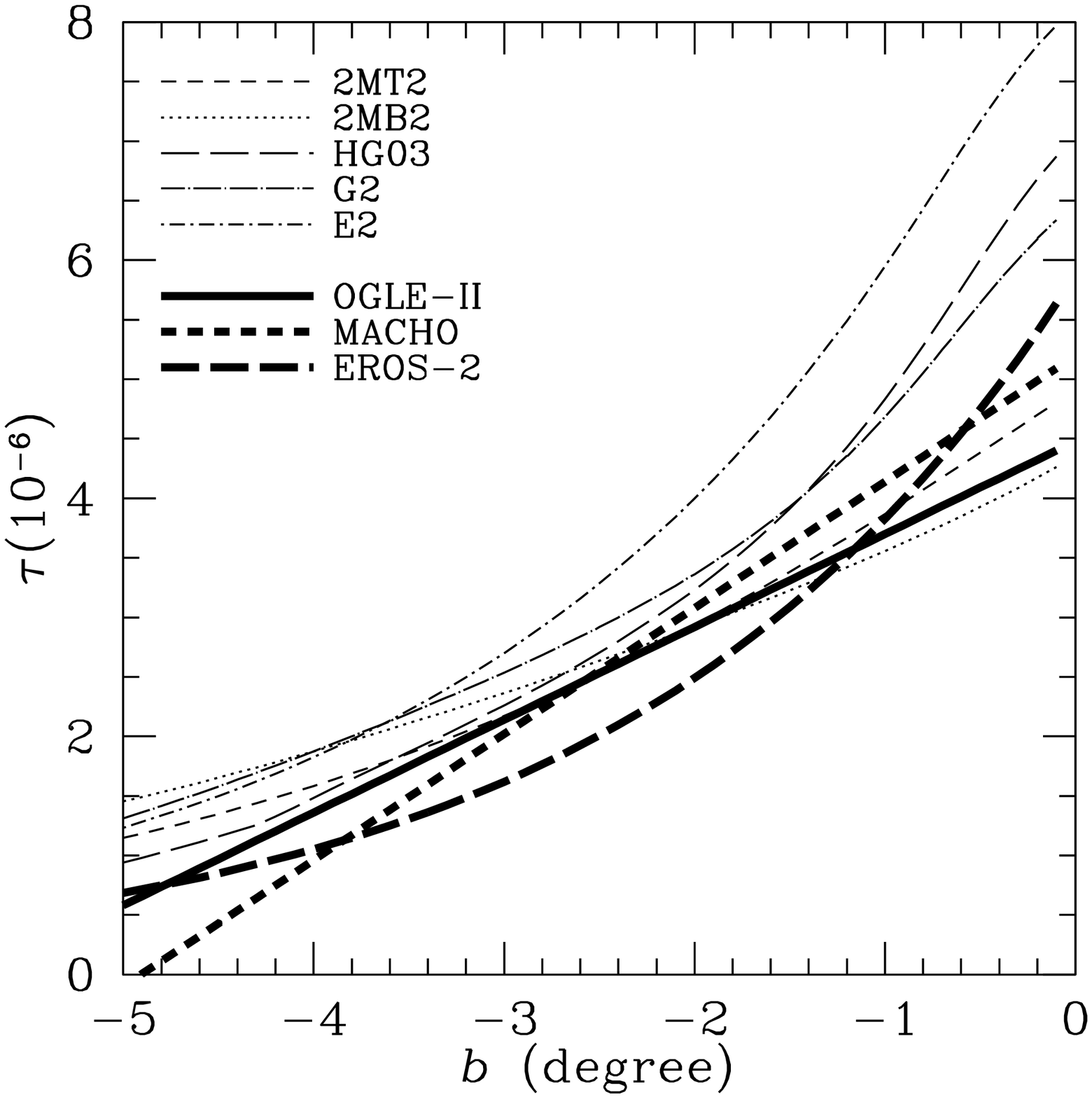}
    \caption{Averaged microlensing optical depths overthe Galactic longitude $l$
    as a function of the Galactic latitude $b$.
    \textit{Left}: Averages of microlensing optical depths measured by OGLE-II,
    MACHO, and EROS-2 are shown with their error bars. OGLE-II, MACHO, and EROS-2
    data are plotted with filled circles, squares, and triangles,
    respectively. Continuous, short dashed, and long dashed lines stand
    for the best-fit curves of OGLE-II, MACHO, and EROS-2, respectively.
    \textit{Right}: Microlensing optical depths toward the Galactic bulge expected from
    models are shown as a function of the Galactic latitude $b$ with
    thin curves. Dashed and dotted curves represent the calculated
    microlensing optical depth resulted from 2MT2 and 2MB2,
    respectively. Long-dashed, dot and long-dashed, and dot and
    short-dashed curves represent HG03, G2, and E2, respectively. For
    comparison, OGLE-II, MACHO, and EROS-2 microlensing experiment
    results are shown with thick continuous, short-dashed, and
    long-dashed curves.}
\end{figure}

\end{document}